# Production of Low-Density Aerogel Nuclear Fuels for Use in Fission Fragment Rockets and Novel Reactor Design

Noah D'Amico [a]*, Sandeep Puri [a], Ian Jones [a], Andrew Gillespie [a], Cuikun Lin [a], Bo Zhao [b], R. V. Duncan [a]

[a] *Center for Emerging Energy Sciences, Department of Physics and Astronomy, Texas Tech University, Lubbock, Texas, USA*
[b] *College of Arts and Sciences Microscopy, Texas Tech University, Lubbock, Texas, USA*
**\* Corresponding Authors:** ndamico@ttu.edu

## Abstract

Graphene hydrogels were created and loaded with uranyl nitrate or thorium nitrate and freeze-dried to produce graphene aerogel nuclear fuels. These aerogels had densities between 0.018-0.035 g/cm$^3$ and consisted of ~7.3± 0.5% uranium/thorium by mass. The ultra-low density of the aerogels allows for high energy ions to escape the fuel particles without depositing all their energy as heat, as is typical in nuclear fuels. Their measured alpha activity was ~16 pCi/mg, which could be enhanced up to ~49 pCi/mg by decreasing the thickness of aerogel samples to allow all alpha particles to escape. Additionally, high energy neutrons were used to induce fission to provide a source of fission fragments from the aerogel fuels. This novel form of nuclear fuel has potential applications in space propulsion such as fission fragment rocket engines, as well as in terrestrial applications for modular reactors, direct conversion methods, and in medical radiotherapeutics.

## 1.0 Introduction

The development of novel nuclear fuels is central to advancing both terrestrial nuclear energy systems and space propulsion technologies.[1–3] One promising approach involves suspending fissionable materials such as natural uranium or thorium within ultralight graphene aerogels to create low-density, porous nuclear fuels.[4] This technique could be applied to fissile isotopes as well, allowing for fission without fast neutrons. The unique structure of graphene aerogels – characterized by extreme surface area, low atomic number, and exceptional thermal stability[5] – provides a medium in which fission fragments (FFs) generated during nuclear reactions can escape the fuel matrix with minimal energy deposition. This sharply contrasts with conventional solid fuels, where almost all of the fission energy is absorbed as heat within the material, necessitating complex cooling systems.

The ability to reduce internal energy deposition opens new avenues for reactor design, including the potential for direct conversion of FF kinetic energy into electricity or thrust. In particular, fission fragment rocket engines (FFREs), long theorized for high-efficiency space propulsion, stand to benefit from such fuel concepts, as escaping FFs can be directed to produce thrust without the thermal and structural limitations of traditional fuels.[6–9] Additionally, the scalability, lightweight nature, and high-temperature resilience of graphene aerogels could enable compact, modular reactor designs for terrestrial applications, including remote power generation and next-generation microreactors.[10,11] In addition, the FF with their full kinetic energy intact may be used to selectively kill malignant tissue and other unwanted tissue that they are placed near within the body of humans and animals, though this idea is highly speculative and carries the constraint of introducing neutrons within the body.

This paper presents a method for creating natural-type uranium and thorium suspensions in graphene aerogels using a hydrothermal synthesis and freeze-drying process. The process resulted in a large relative density of fissionable isotopes ($^{238}$U or $^{232}$Th) within the aerogel that may act as an alpha particle source. Activities were measured using CR-39 plastic nuclear track detectors (PNTDs). Upon exposure to 14 MeV neutrons from D-T fusion, the aerogel fuel produced observable FF tracks in CR-39 detectors. Automated characterization and counting of alpha and FF tracks were performed using an artificial intelligence-based object recognition program to ensure high measurement accuracy and reproducibility.

## 2.0 Experiment

### 2.1 Aerogel Synthesis and Irradiation

Graphene oxide (GO) aerogels were synthesized following the procedure described by S.T. Nguyen et al. (2012) with minor modifications.[12] GO aerogel was chosen as the fuel matrix due to ease of manufacturing, greater thermal conductivity, and lower electron density that allows for greater penetration depth of energetic ions compared to silica aerogels. Powdered graphite was oxidized by reacting it with sulfuric acid, sodium nitrate, and potassium permanganate in an ice bath, producing graphite oxide. The resulting material was sonicated for 48 hours to exfoliate the graphite oxide into graphene oxide sheets. To form a graphene oxide hydrogel, 100 mL of this graphene oxide suspension was transferred into a hydrothermal chamber and heated at 120 °C for 20 hours. At this stage, the hydrogels were submerged in 10 mL of 0.02 M natural-type $UO_2(NO_3)_2$ and $Th(NO_3)_4$ solutions to dope the hydrogel with radioisotopes (this doping step was omitted for control samples). We believe that any water-soluble solution containing the desired radioisotopes would be suitable for doping. The doped hydrogels were allowed to adsorb the isotopes for 24 hours before undergoing freeze-drying to yield the final graphene aerogel. The aerogels were then characterized by measuring mass, volume, and alpha particle and gamma emission from the radioisotope decay.

The amount of uranium loaded into the aerogel was measured first by gamma spectrometry. Known masses of $UO_2(NO_3)_2 \cdot 6H_2O$ crystal from 9 mg to 196 mg were measured using a single crystal intrinsic germanium gamma spectrometer[13] for 30 minutes each. A calibration curve of counts vs mass of $UO_2(NO_3)_2 \cdot 6H_2O$ was obtained. All pieces of uranium-loaded aerogel (78mg total) were then measured and fit to this curve. This approach gave a very low signal-to-noise ratio due to the low activity of the sample and reinforced our confidence in CR-39's ability to provide accurate measurement in low signal experiments.

To achieve better signal-to-noise ratio in measuring activity (and therefore uranium content), three pieces of 0.25 to 0.4 cm³ aerogel fuel, totaling 1 cm³ and 22 mg, were placed directly on top of a 3 cm by 2 cm piece of CR-39 (manufactured by BlankSlate Innovation[14]) inside of a covered petri dish, reducing radiation from outside sources such as radon. Note that this measurement was conducted in air, not vacuum, so additional attenuation of alpha particles occurred. These conditions were held for a 48-hour period to determine the alpha activity of the aerogel. Following these exposures, the CR-39 pieces were etched in an 80 °C, 6.25 M NaOH bath for 1 hour with a stir rod at a constant stirring rate of 300 rpm to ensure even etching. The CR-39 was then analyzed under an optical microscope and large area mapping was done using a Zeiss Crossbeam 540 SEM. Control samples in this exposure environment exhibited no alpha particle tracks.

Monte Carlo n-Particle (MCNP®) transport simulations[15] were performed to estimate the solid angle subtended by the CR-39 PNTDs. The MCNP® geometries and source conditions were designed to closely match the experimental parameters. The aerogel pieces were treated as rectangular cuboid volume sources of alpha particles placed on top of a CR-39 detector. Calipers were used to measure the dimensions of the three aerogel pieces to obtain maximum and minimum length in each direction. Emission probabilities were adjusted to match the mass of each aerogel piece. Simulations simply counted the number of particles that escaped the aerogel and the number of those that entered the CR-39 detector. The solid angle simulations were repeated for each set of maximum and minimum dimensions. When simulations involved vacuum in the intermediate space, the portion of the solid angle subtended was found to be between 29.9% and 35.2%. When simulations involved air in the intermediate space, the portion of the solid angle subtended was found to be between 28.5% and 33.5%. These values were used in calculating the aerogel activity and uncertainty.

Similarly, a 0.25 cm³, 8.8 mg, piece of aerogel fuel with 0.5 cm² surface area facing the neutron generator was placed on top of a piece of CR-39 taped directly to the target plane of the P383 neutron generator.[16] Running at 120 keV and 70 µA, this gave an expected flux of $6.7 \times 10^5$ n/s on the aerogel, or $4.8 \times 10^9$ total neutrons incident over the irradiation time of two hours. This figure is based on manufacturer specifications described in the user manual of the P383. Neutrons passed through CR-39 detector, leaving typical high energy neutron background tracks. Upon reaching the aerogel, the neutrons reacted with the $^{238}U$ or $^{232}Th$ to create FFs, which escaped the aerogel matrix to leave a larger, darker track within the CR-39. This setup was repeated with undoped GO aerogel as well. The irradiation setup is shown in *Figure 1*. The FF tracks and 14

MeV neutron tracks were visualized under an optical microscope and SEM after applying the aforementioned etching process.

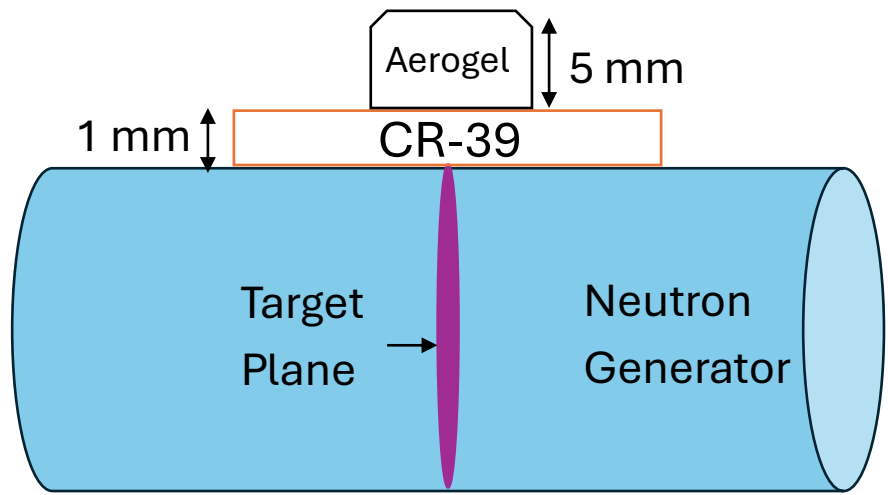

**Figure 1:** A simplified depiction of the neutron generator, CR-39 detector, and fissionable GO aerogel.

### 2.2 SEM

SEM was used to create large area maps to analyze emergent phenomena using a Zeiss Crossbeam 540 with Atlas 5 large area mapping imaging system.[17] This SEM is a powerful tool for advanced materials research capable of precise SEM imaging to enable high-resolution surface imaging and site-specific sample preparation. With sub-nanometer resolution, it can deliver high-contrast imaging of surfaces and subsurface structures. Its high throughput capabilities were combined with advanced automation methods to optimize fast data acquisition with our automated workflow method. The images were collected under 5 kV accelerating voltage with a SE detector. The large area mapping parameters were set to a resolution of 200 nm with 6.4 µs dwell time, 600 µm by 600 µm tile size, and 10% overlap. It should be noted that this overlap setting is purely for the automated construction of the mosaic and does not influence the source images. Therefore, the total particle count is not affected by the 10% overlap.

Images were captured robotically, spanning the entire surface, and stitched together using Mosaic software.[17] The AI system was trained to characterize particle tracks left on the surface of the CR-39 large area map based on diameter, grayscale contrast, solidity and eccentricity. The images that constitute the large area map were analyzed by the AI individually and later stitched together to recreate a post-processing large area map.

### 2.3 Machine Learning Object Detection Analysis

MATLAB machine learning and object detection packages were used to develop an AI object detection program for use on CR-39. For automated analysis of particle tracks, a transfer-learning-based Region-based Convolutional Neural Network (R-CNN) was employed.[18,22–24] The

R-CNN algorithm utilizes a pre-trained deep residual neural network (ResNet-101) backbone to identify and classify regions of interest, specifically tailored to detect the small, elliptical features characteristic of alpha particle tracks. To improve detection accuracy and computational efficiency, a Canny edge detection algorithm was applied to highlight particle tracks in the SEM images. The CR-39 track detection program is described in more detail in D'Amico et al.'s 2025 paper.[25]

The FFs were distinguished using a combination of pre- and postprocessing of the AI detections. First, Canny parameters were applied to give a general neutron background detection: low thresholds and blur (low threshold = 0.06, high threshold = 0.2, sigma = 4) to detect all neutron background and FF tracks, then a filter rejecting bounding boxes above 1000 pixel area was applied to remove FFs and abnormally large detections that usually were caused by overlapping tracks.

To reliably identify FF tracks while reducing false positives from surface defects and overlapping tracks, we employed a seeding and reconstruction approach for track detection. In this process, an initial set of candidate regions ("seeds") is identified using edge-detection and intensity-based criteria that highlight the sharply etched, elongated morphology characteristic of FF tracks. These seeds serve as anchor points from which track boundaries are reconstructed by following local gradients and edge continuity. After extensive trial and error, combined with human-graded accuracy assessments, we determined a set of parameters that reliably isolated FFs while rejecting background and noise. Specifically, the image was segmented using intensity thresholds (seedThresh = 0.16, maskThresh = 0.40) to initialize seed and mask regions, followed by size filtering (bwareaopen with a minimum area of 30 pixels for general features and 600 pixels for candidate FFs) and morphological reconstruction. Connected components were then characterized by area, perimeter, solidity, circularity, and mean intensity. Final selection criteria required features to satisfy (area < 2000 pixels$^2$, circularity > 0.25, and 0.6 < solidity < 0.98), which eliminated most neutron background and surface defects while preserving true FFs.[26–29]

This reconstruction step not only improves track fidelity but also reduces overcounting from fragmented detections. Surface imperfections tend to generate irregular, disconnected features that fail to propagate beyond the seeding stage, whereas real FFs consistently reconstruct into elongated, high-aspect-ratio blobs. The seeding and reconstruction process created a calculated detection precision of 94%, recall of 89%, and overall F1-score of 92%. These scores were determined via human validation of images (300 images containing 1.6 FFs on average) from the samples discussed in this paper. The images with bounding boxes for neutron background and FFs were then stitched back into a large area map for visualization.

For analysis of the alpha decay from aerogel fuels, a system with two sets of parameters was also used. The first used simple Canny parameters with low threshold and low blur (low threshold = 0.12, high threshold = 0.17, sigma = 1), which detected all alpha particle tracks regardless of degree of moderation. The second created dark pixel masks, masking any locations with pixels darker than 80 on the grayscale (0 to 255 from dark to light), and then applied a gaussian blur with sigma = 10 to the remainder of the image. This process keeps all dark, deep

(and therefore high energy) alpha tracks, and blurs out the more shallow and light edges to remove low energy alpha tracks. Slightly higher Canny thresholds and blurs were used in this detection (low threshold = 0.15, high threshold = 0.35, sigma = 2). The results and reasoning for using this scheme are discussed in the ***Results*** section and ***Figure 3***.

## 3.0 Results

After freeze-drying, the aerogel pieces ranged in shape and size between 0.25 and 4 cm$^3$. The large variations in size came from handling the hydrogel before and during the freeze-drying process, as the hydrogel is very fragile and can easily be split by tweezers. We believe this process can be regulated better for more precise control over sample shape and size, but this will be reserved for future works. Control samples without Th or $UO_2$ loading had densities ranging from 0.011 to 0.025 g/cm$^3$, while samples with Th or $UO_2$ loading had densities ranging from 0.018 to 0.035 g/cm$^3$. The following results are presented for $UO_2$-loaded samples only for brevity, but the experimental procedures yield very similar results for Th-loaded aerogel.

The calibration curve generated for the $UO_2(NO_3)_2 \cdot 6H_2O$ gamma activity is shown in ***Figure 2***. The 78 mg sample of uranium-loaded aerogel gave off 154 counts in 30 minutes, but the background signal was 132 counts in 30 minutes, leading to very low signal-to-noise ratio. 154 counts corresponds to a mass of 18.2 ± 10.3 mg of $UO_2(NO_3)_2 \cdot 6H_2O$, or 8.6 ± 4.8 mg of natural uranium. Therefore, the aerogel has an average loading density of 11 ± 6% uranium by mass, and an activity of 79 ± 45 pCi/mg aerogel. It is clear that this error is too high to use the figure as conclusive proof of uranium loading density, and CR-39 provides a more accurate determination of uranium content in low signal environments.

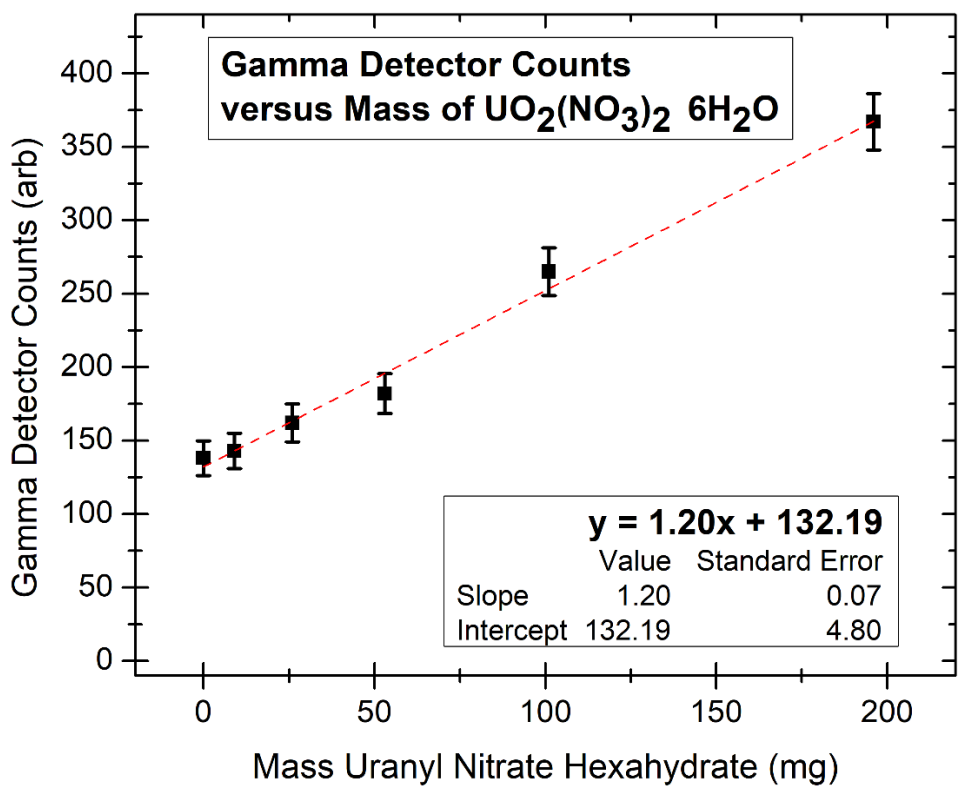

**Figure 2:** Calibration curve of gamma detector counts from known masses of $UO_2(NO_3)_2 \cdot 6H_2O$ with collection times of 30 minutes per data point. The linear fit equation ($R^2$ = 0.9911) was used to determine the mass of uranium embedded in the aerogel after a similar gamma measurement yielded 154 detector counts.

The alpha decay activity of the aerogel fuels was determined by placing the aerogel directly on top of a piece of CR-39 for 48 hours. After etching, a large area mapping (LAM) of the CR-39 surface was created using the SEM. The AI track counting program was applied twice with different parameters discussed in the *Experiment* section: once to detect only high energy alphas, and once to detect all visible alpha particle tracks, regardless of moderation. Note the dark features that mark deep penetration in the high energy alpha image – many lighter tracks corresponding to low-energy alpha particles are visible as well, but these were only included in the "full energy spectrum" detection which led to the total activity calculation. These lower energy alphas are expected throughout the CR-39 sample, as the ~4.25 MeV alpha particles from $^{238}U$ decay have a stopping distance of ~3.5 cm in air, according to SRIM (Stopping Range in Matter Code). There is still correlation between track location and aerogel in the low energy alpha heat map, but the edges surrounding the aerogel also have high numbers of tracks. Sample detection images and the resulting track density heat maps are included in *Figure 3.* Using the simulated subtended solid angle of 32.5% ± 2.6%, the total track count of 764,000 ± 65,000 over 48 hours corresponds to an activity of 368 ± 34 pCi, or 16.7 ± 1.5 pCi/mg. Simulations using the maximum and minimum dimensions of the aerogel pieces were conducted, and it was found that about 60% of the emitted alpha particles do not escape the aerogel pieces of this size, and another 2% may be lost while traveling in air to the edges of the CR-39. It follows that decreasing aerogel thickness would increase the alpha activity per mass of aerogel until the total thickness of each dimension drops below the penetration depth of 4.25 MeV alpha particles (1.8mm), though this was not attempted experimentally. Calculations of aerogel samples with dimensions below the penetration depth of alpha particles suggest that the true alpha activity of the aerogel is 49 ± 4.5 pCi/mg. This gives a uranium loading density of 7 ± 0.7% by mass. It is worth mentioning that no discernable alpha tracks were found in the control (without uranium loading) aerogel sample.

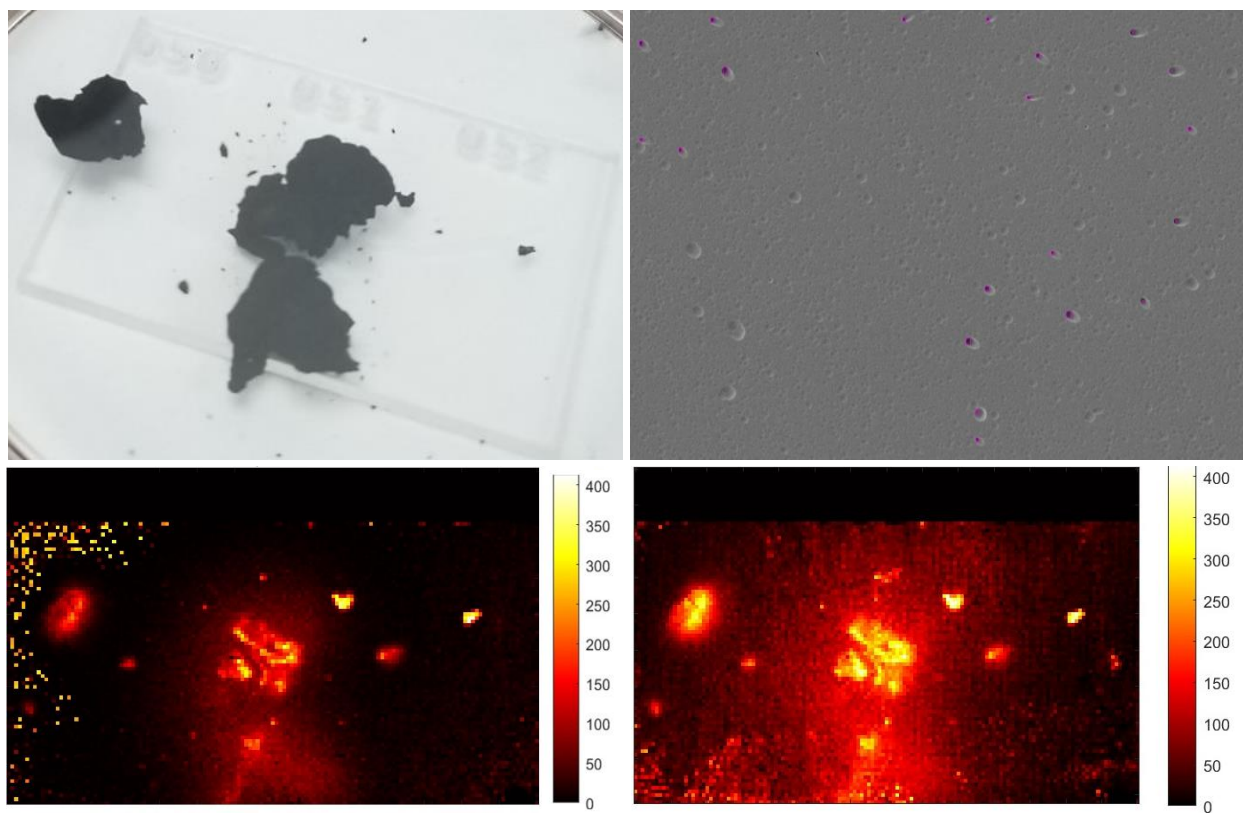

**Figure 3:** (**top left**) Pieces of uranium-doped GO aerogel on CR-39, left for 48-hour irradiation. (**top right**) Sample SEM image with AI detection (magenta boxes) of only high-energy alpha particles. (**bottom left**) Heat map of high-energy alpha particle tracks on the CR-39 – note the direct correlation between the aerogel placement on CR-39 during exposure and the density of high-energy alphas. (**bottom right**) Heat map of low-energy alpha particle tracks on the CR-39.

From $4.8\times10^9$ total 14 MeV neutrons incident, 5,590 ± 475 FF tracks were recorded in the CR-39. Simulations using the minimum and maximum dimensions of the aerogel found that the CR-39 subtended 29.9% and 35.2%, respectively, of the solid angle from the aerogel. It is estimated that the total number of FFs that exited the aerogel was 18,032 ± 2000. Using SRIM simulations, the stopping distance of FFs in aerogel (simulated as 0.02 g/cm³ $C_2O$) was found to be 1 to 1.2 mm, depending on the atomic mass and energy distribution. Accounting for the fact that our sample size was larger than the stopping distance, we find that the 18,032 tracks originated from only the outer ~40% of the sample volume. The true figure for FFs created is around 45,000 ± 4700. Back-of-the-envelope calculation of uranium atom number necessary to create this flux (using $N_{reactions} = \Phi*\sigma*N_{atoms}$), leads to $N_{atoms}$ = 1.8 (±0.2) x $10^{18}$, or 0.7 ± 0.1 mg U in the 8.8 mg aerogel sample. This implies a uranium loading of 8 ± 1.2% by mass.

Using SRIM, the stopping distance of FFs in air was found to be 2 to 3 cm. This was evident in the experiment, as FF tracks were found in lower abundance in the CR-39 over 1 cm away from where the aerogel lay. **Figure 4** shows the heat map of FF tracks in the 2 by 1 cm CR-39 piece, along with a detection image of neutron background and fission fragments. The AI track detection/characterization program detected 104 FF tracks in the control (no uranium loading), but human inspection determined that these were all false positive detections. This is expected, as the precision (1 – false positive rate) was only 94%.

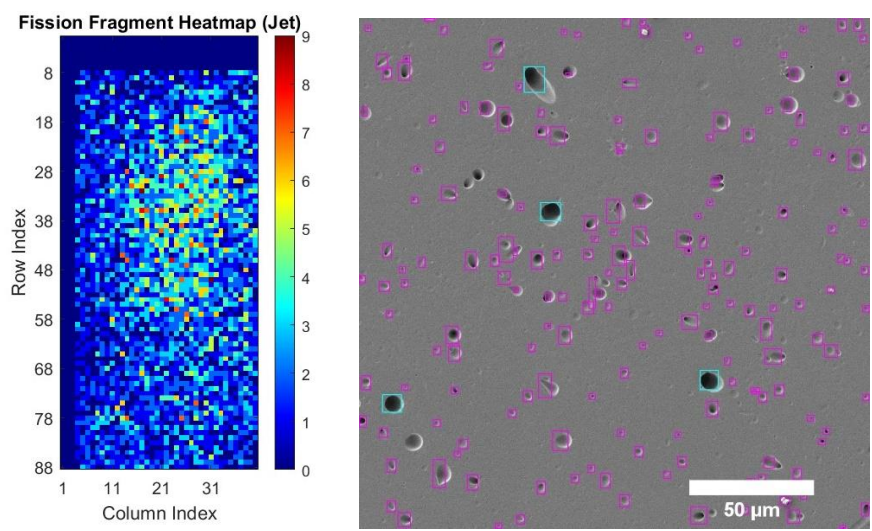

**Figure 4: (left)** Heat map of FF tracks on a 2 by 1 cm piece of CR-39 after 14 MeV neutron irradiation of uranium-loaded GO – note that the position and shape of the aerogel is obvious from the density of FF tracks in the CR-39. (**right**) Sample image of FF detection (blue) with 14 MeV neutron background (pink).

**4.0 Discussion**

As stated in the results section, the large error present in the gamma spectrometry measurement of uranium content in aerogel points to the inability of many traditional nuclear detection techniques to provide precise measurements in low signal-to-noise environments. CR-39 has the exact opposite effect when measuring energetic ions. Experiments can typically be adequately isolated from external charged particle sources to ensure extremely low background noise on CR-39 detectors (less than 10 tracks per $cm^2$). As long as experiments are allowed to run for sufficient time to attain significant signal, there is very little inherent error. The largest source of error in CR-39 measurement comes from the post-processing, which is shown to be below 10% in this paper.

This paper presents a method for synthesis of fissionable aerogel fuels that can be utilized in applications that require high energy ions to escape the fuel matrix. One such application is the Fission Fragment Rocket Engine (FFRE), in which FFs are channeled to provide thrust for a rocket in deep space travel. This rocket engine provides enormous $I_{sp}$ values (>500,000) with low thrust and can be integrated directly with Nuclear Thermal Rockets (NTRs), allowing for a complementary high thrust, low $I_{sp}$ stage for escaping orbit. For this task, aerogel sheets of ~20-micron thickness could be used to eject FFs with only ~1% energy loss on average, or ~200-micron thickness could be used if ~10% energy loss was allowable. Assuming the surface area was made relatively large by increasing the other dimensions of the sheets (10 cm by 10 cm), the radiative cooling of these sheets would keep them well below graphene aerogel disintegration temperature.

Fissionable aerogel fuels also allow for the implementation of direct conversion methods that will convert the high energy (1 MeV to 100 MeV) ions of nuclear reactions into electricity. While highly efficient methods for direct conversion have not been discovered yet, many sources on the order of 50 mW/g power density exist, and federal initiatives are funding new research to increase to W/g levels of power generation.[30–32] While thin films (~1 micron) of solid fissionable material deposited onto substrates are currently being used for such tasks, aerogel fuel allows for greater percentage of particle escape and higher escape energies while removing the need for a substrate and allowing a multitude of fixture configurations.

As a speculative concept, an aerogel-based fission source placed at a tumor site and irradiated with a fast-neutron beam could, in principle, generate very-high–LET particles (FFs) in situ that deposit enormous energy over micron-scale ranges. This spatially concentrated energy might be highly effective against radioresistant, hypoxic, or otherwise hard-to-treat tumor cells in a way analogous to existing high-LET modalities (heavy ions, alpha emitters) and to neutron-capture therapies that rely on localized nuclear reactions. However, the downsides and barriers are substantial. Triggering fission produces not only energetic charged fragments but also prompt and delayed neutrons, gamma rays, and a suite of short- and long-lived fission products; those emissions complicate shielding, increase whole-body and bystander exposure, and could activate surrounding materials or create systemic contamination if radioactive particles escape. While this is an intriguing medical application, the radiological hazards and practical constraints mean it remains, at best, a high-risk research idea requiring extensive preclinical study before any clinical consideration. [33,34]

## 5.0 Conclusion

Graphene hydrogels were created and loaded with uranyl nitrate or thorium nitrate and freeze-dried to produce graphene aerogel nuclear fuels. This produced aerogels with ~7.3 ± 0.5% uranium/thorium loading by mass. The ultra-low density of the aerogels allows for alpha particles and fast fission fragments to escape the fuel particles without depositing all their energy as heat. Their measured alpha activity was ~16 pCi/mg, but total isotope loading suggests that an activity up to ~49 pCi/mg could be attained by decreasing the thickness of aerogel pieces though this enhancement is beyond the scope of the presented experiment. Additionally, high energy neutrons were used to induce fission of the radioisotopes and produce fission fragments to directly demonstrate the escape of fission fragments from the aerogel fuels. This novel form of nuclear fuel has potential applications in space propulsion such as fission fragment rocket engines, in new medical radiotherapeutics, as well as in terrestrial applications for modular reactors and direct conversion methods.

## 5.0 Conflict of Interest

The authors declare that the research was conducted in the absence of any commercial or financial relationships that could be construed as a potential conflict of interest.

## 6.0 Author Contributions

All authors contributed to the conception and design of the research reported here. ND wrote the first draft of the manuscript. All authors contributed to manuscript revision, read, and approved the submitted version.

## 7.0 Funding

This work was supported by the Department of Energy award No. DE-AR0001736, the Texas Research Incentive Program and Texas Tech University.

## 8.0 References

1.  Peakman, A. & Lindley, B. A review of nuclear electric fission space reactor technologies for achieving high-power output and operating with HALEU fuel. *Prog. Nucl. Energy* **163**, 104815 (2023).

2.  Zhan, L., Bo, Y., Lin, T. & Fan, Z. Development and outlook of advanced nuclear energy technology. *Energy Strategy Rev.* **34**, 100630 (2021).

3.  Andersson, D. A., Stanek, C. R., Matthews, C. & Uberuaga, B. P. The past, present, and future of nuclear fuel. *MRS Bull.* **48**, 1154–1162 (2023).

4.  Gorgolis, G. *et al.* Graphene aerogels as efficient adsorbers of water pollutants and their effect of drying methods. *Sci. Rep.* **14**, 8029 (2024).

5.  Nassar, G., Daou, E., Najjar, R., Bassil, M. & Habchi, R. A review on the current research on graphene-based aerogels and their applications. *Carbon Trends* **4**, 100065 (2021).

6.  Weed, R. Aerogel Core Fission Fragment Rocket Engine - NASA. https://www.nasa.gov/general/aerogel-core-fission-fragment-rocket-engine/.

7.  Puri, S., Gillespie, A., Duncan, R. V. & Lin, C. Alpha particle detection in high magnetic fields with applications in designing fission fragment rocket engine | Scientific Reports. https://www.nature.com/articles/s41598-025-07556-8.

8. Gahl, J., Duncan, R. V., Lin, C. & Gillespie, A. Frontiers | The fission fragment rocket engine for Mars fast transit. https://www.frontiersin.org/journals/space-technologies/articles/10.3389/frspt.2023.1191300/full.

9. Werka, R. Final Report: Concept Assessment of a Fission Fragment Rocket Engine (FFRE) Propelled Spacecraft.

10. Black, G., Shropshire, D., Araújo, K. & van Heek, A. Prospects for Nuclear Microreactors: A Review of the Technology, Economics, and Regulatory Considerations. *Nucl. Technol.* **209**, S1–S20 (2023).

11. Bryan, H. C., Jesse, K. W., Miller, C. A. & Browning, J. M. Remote nuclear microreactors: a preliminary economic evaluation of digital twins and centralized offsite control. *Front. Nucl. Eng.* **2**, (2023).

12. Nguyen, S. T. *et al.* Morphology control and thermal stability of binderless-graphene aerogels from graphite for energy storage applications. *Colloids Surf. Physicochem. Eng. Asp.* **414**, 352–358 (2012).

13. High-Precision Spectroscopy with PGT HPGe. *Maximus Energy* https://maximus.energy/index.php/2023/02/23/high-precision-spectroscopy-with-pgt-hpge/ (2023).

14. Blank Slate Innovation | Advancing science one project at time. https://blankslateinnovation.com/.

15. Kulesza, J. A. MCNP® Code Version 6.3.0 Theory & User Manual. (2022).

16. P 383 Neutron Generator Each | Request for Quote | Thermo Scientific™ | thermofisher.com. https://www.thermofisher.com/order/catalog/product/153993-G1.

17. Rauscher, M. ZEISS Atlas 5 Large Area Imaging with High Throughput. (2014).

18. Kordemir, M., Cevik, K. K. & Bozkurt, A. A mask R-CNN approach for detection and classification of brain tumours from MR images. *Comput. Methods Biomech. Biomed. Eng. Imaging Vis.* **11**, 2301391 (2024).

22. Sun, P. *et al.* Sparse R-CNN: An End-to-End Framework for Object Detection. *IEEE Trans Pattern Anal Mach Intell* **45**, 15650–15664 (2023).

23. Puchaicela-Lozano, M. S. *et al.* Deep Learning for Glaucoma Detection: R-CNN ResNet-50 and Image Segmentation. *J. Adv. Inf. Technol.* **14**, 1186–1197 (2023).

24. Qin, H. *et al.* An Improved Faster R-CNN Method for Landslide Detection in Remote Sensing Images. *J. Geovisualization Spat. Anal.* **8**, (2023).

25. D'Amico, N. *et al.* AI-Based Large-Area Nuclear Particle Track Analysis System. *Nucl. Eng. Technol.* **57**, 103738 (2025).

26. Xu, Y. *et al.* Advances in Medical Image Segmentation: A Comprehensive Review of Traditional, Deep Learning and Hybrid Approaches. *Bioengineering* **11**, 1034 (2024).

27. Long, J., Liu, Y. & Zhang, K. Interactive image segmentation combining global seeding and sparse local reconstruction | Pattern Analysis and Applications. https://link.springer.com/article/10.1007/s10044-025-01432-x.

28. Jiang, J.-A., Chuang, C.-L., Lu, Y.-L. & Fahn, C.-S. Mathematical-morphology-based edge detectors for detection of thin edges in low-contrast regions. *IET Image Process.* **1**, 269–277 (2007).

29. Santos, E. M. D. S. & Marcal, A. R. S. Segmentation of Microscopic Images for Pollen Grains Detection | 8th International Conference of Pattern Recognition Systems (ICPRS 2017). https://digital-library.theiet.org/doi/abs/10.1049/cp.2017.0169.

30. Rads to Watts | DARPA. https://www.darpa.mil/research/programs/rads-watts.

31. Stefanishin, D. I., Игоревич, С. Д., Davydova, S. O. & Олеговна, Д. С. DIRECT CONVERSION OF RADIATION INTO ELECTRICITY. *Вестник Молодых Учёных И Специалистов Самарского Университета* 119–124 (2021).

32. Spencer, M. G. & Alam, T. High power direct energy conversion by nuclear batteries. *Appl. Phys. Rev.* **6**, 031305 (2019).

33. Taylor, G. N. *et al.* Fission Fragment Relative Biological Effectiveness for Liver Tumor Induction. *Health Phys.* **69**, 269 (1995).

34. Gordon, K., Gulidov, I., Fatkhudinov, T., Koryakin, S. & Kaprin, A. Fast and Furious: Fast Neutron Therapy in Cancer Treatment. *Int. J. Part. Ther.* **9**, 59–69 (2022).